\documentclass[aps,pra,superscriptaddress,showpacs,showkeys,a4paper,10pt,notitlepage,twocolumn]{revtex4-2}
\usepackage[utf8]{inputenc}
\usepackage[english]{babel}
\usepackage{amsmath}
\usepackage{amssymb}
\usepackage{amsfonts}
\usepackage{graphicx}
\usepackage{braket}
\usepackage{diagbox}
\usepackage{colortbl}
\usepackage{cancel}
\usepackage{upgreek}
\usepackage{mathtools}
\usepackage[T1]{fontenc}

\usepackage{xcolor}
\definecolor{link_blue}{RGB}{52,46,157}

\usepackage{longtable}
\usepackage[colorlinks,citecolor=link_blue,linkcolor=link_blue,urlcolor=link_blue]{hyperref}
\usepackage[tiny,center,uppercase]{titlesec}

\usepackage{tikz,xcolor}
\definecolor{lime}{HTML}{A6CE39}
\DeclareRobustCommand{\orcidicon}{%
 \begin{tikzpicture}
    \draw[lime, fill=lime] (0,0)
    circle [radius=0.14]
    node[white] {{\fontfamily{qag}\selectfont \tiny ID}};
    \draw[white, fill=white] (-0.0625,0.095)
    circle [radius=0.007];
    \end{tikzpicture}
    \hspace{-2mm}
}

\renewcommand{\vec}{\boldsymbol}

\begin{document}

\title{X-ray resonance therapy  with parametric X-ray radiation (PXR)
for sulfur-containing tumor tissues }

\newcommand{\orcidSan}
{\href{https://orcid.org/0000-0003-2919-5414}{\orcidicon}}

\newcommand{\orcidODS}
{\href{https://orcid.org/0000-0002-2875-0140}{\orcidicon}}

\newcommand{\orcidLeonau}
{\href{https://orcid.org/0000-0002-4830-6856}{\orcidicon}}

\newcommand{\orcidIDF}
{\href{https://orcid.org/0000-0003-0476-8634}{\orcidicon}}

\author{N.\ Q.\ San\orcidSan}
\email[Corresponding author: ]{nguyenquangsan@hueuni.edu.vn}
\affiliation{School of Engineering and Technology - Hue University, Hue, Vietnam}

\author{O.\ D.\ Skoromnik\orcidODS}
\affiliation{Currently without university affiliation}

\author{A.\ U.\ Leonau\orcidLeonau}
\affiliation{Currently without university affiliation}

\author{V.\ Q.\ Nha}
\affiliation{School of Engineering and Technology - Hue University, Hue, Vietnam}

\author{I.\ D.\ Feranchuk\orcidIDF}
\affiliation{Atomicus GmbH Amalienbadstr. 41C, 76227 Karlsruhe, Germany}
\affiliation{Belarusian State University, 4 Nezavisimosty Ave., 220030 Minsk, Belarus}

\begin{abstract}
This study investigates the feasibility of using parametric X-ray radiation (PXR) for the selective therapy of superficial tumor tissues with elevated sulfur concentrations. The therapeutic mechanism is based on the resonant ionization of sulfur atoms at the $K$-edge ($\hbar\omega_K \approx 2.4$ keV), a frequency range where PXR exhibits narrow spectral distribution and high tunability. We calculate the photon flux density and total absorbed doses required to achieve irreversible damage in tumor cells while minimizing impact on healthy tissues. Our results demonstrate that a therapeutic effect is attainable with an absorbed dose of approximately $1.3$ Gy, significantly lower than the doses required by conventional X-ray tubes. Furthermore, the study shows that PXR-based sources can operate at electron energies of $20$--$40$ MeV. This energy range is compatible with compact medical accelerators, offering a practical advantage over large-scale synchrotron facilities for clinical deployment in standard hospital environments.
\end{abstract}



\maketitle

\section{Motivation}
\label{sec:motivation}
Currently, many medical technologies for diagnosis and treatment are based on the use of physical phenomena \cite{sandoghdar2024essay}. One of the important directions in this field is radiation therapy for cancer, which involves the destruction of tumor cells that form the pathological area. An essential requirement of radiation therapy is that the impact on healthy cells should be minimized as much as possible \cite{kudriashov2008radiation}. In standard X-ray sources based on bremsstrahlung radiation from electrons, the photon beam has a wide frequency spectrum and covers a large area of the irradiated object, so the resulting radiation dose is not used efficiently. To reduce the total irradiation dose, a promising approach involves using synchrotron radiation as a source, which has a narrow frequency spectrum and small angular divergence \cite{labriet2018significant}. However, the application of this method requires relatively large accelerators with electron energies on the order of 1 GeV.

In this paper, we consider the possibility of using parametric X-ray radiation (PXR) \cite{PXR_Book_Feranchuk} as a source of radiation for tumor treatment. This radiation generation mechanism makes it possible to produce a photon beam with characteristics similar to those of synchrotron radiation, but at electron energies on the order of $20$--$40$ MeV, which are typical for modern medical accelerators \cite{Podgorsak2014}. The use of a compact radiation source based on PXR \cite{balanov2024coherentinteractionsfreeelectrons} could significantly improve the effectiveness of radiation therapy for certain types of cancer.

In order to characterize the irradiation efficiency the following quantity can be determined
\begin{align}
  K(D) = \frac{N_{i}(D)}{N_{h}(D)}, \label{eq:1}
\end{align}
where $N_{i}(D)$ is the number of tumor cells, which were damaged during the irradiation of a patient with the radiation dose $D$ and $N_{h}(D)$ is the number of the corresponding damaged healthy cells, during irradiation with the same dose $D$. In most cases the atomic structures of the tumor and healthy cells are very similar and the value $K(D) \leq 1$. As larger the value $K(D)$ for some type of the radiation is as it is more effective for the therapy
\cite{Kuzin1991}.

Another factor in the efficiency of radiation therapy is the number of tumor cells $N_{i}(D)$ reaching a critical value $N_{i}^{\mathrm{cr}}(D)$, at which irreversible damage to the tumor tissue occurs. Different types of radiation therapy are successful at different radiation doses $D_{f}$. Therefore, when comparing two types of radiation therapy, $f_{1}$ and $f_{2}$, we can introduce the parameter
\begin{align}
  \xi_{1,2} = \frac{D_{f_{1}}}{D_{f_{2}}}. \label{eq:2}
\end{align}
For a given tumor type, the optimal radiation therapy is chosen such that the parameter $K$ is maximized, while $\xi_{1,2}$ is minimized.

The basis for the proposed project is the experimental observation that in certain tumor types, the accumulation of sulfur-containing glutathione \cite{gamcsik2012glutathione,huang2001mechanism} and methionine \cite{CELLARIER2003489} molecules leads to an increase in the concentration of sulfur atoms, which is approximately three times higher than in healthy tissues. Therefore, if a type of radiation is used that specifically targets sulfur atoms, the parameter
\begin{align}
  K_{S}(D) \approx 3. \label{eq:3}
\end{align}

Sulfur has particularly rich biochemistry and plays a number of important roles in structure, catalysis, and metabolism in all organisms \cite{czapla2013chemical}. Therefore, we can assume that the action of radiation on tumor cells through the ionization of sulfur atoms initiates a cascade of secondary processes, such as characteristic radiation, Auger recombination, and collision ionization, which lead to the breakage of molecular bonds. This triggers the chemical and biological stages of radiation action and, as a consequence, cell death \cite{Kudrashov2004}. According to experimental observations \cite{Kuzin1991}, the lethal power of the absorbed irradiation dose can be estimated as
\begin{align}
 \frac{D_f}{\Delta t} \approx 10  \ \frac{Gy}{min},  \label{eq:4}
\end{align}
where $Gy = J/kg$ is the unit of the integral absorbed dose.

\section{Qualitative analysis}
\label{sec:qualitative}
In order to estimate the efficiency of this kind of the radiation therapy we will use the ionization cross section of the sulfur atom. In the case of electromagnetic radiation, this quantity has a resonant character \cite{Landau2002}, and if we neglect the small splitting of the energy levels in the innermost atomic shells, it can be represented as
\begin{align}
  \sigma_{S}(\omega) &= \sigma_{r}(\omega) + \sigma_{nr}, \label{eq:4}
  \\
  \sigma_{r}(\omega) &\approx \sigma_{0}
  \frac{\Gamma_{\omega}^{2}}{(\omega - \omega_{K})^{2} +
  \Gamma_{\omega}^{2}}. \label{eq:5}
\end{align}
Here $\sigma_{r}(\omega)$ is the resonance part of the cross section, which has a sharp peak near the $K$-edge, with the frequency $\omega_{K}$ of the absorption line of the sulfur atom; $\Gamma_{\omega}$ determines the width of the peak, and $\sigma_{nr}$ is the nonresonant part of the cross section, which is almost constant over a wide range of frequencies.

For the case of the sulfur atom, these quantities are \cite{mirkin1964handbook}:
\begin{align}
  \sigma_{0}
  &\approx 9.2 \times 10^{-20}\,\mathrm{cm}^{2}, \quad
  \frac{\sigma_{0}}{\sigma_{nr}} = 11.4, \label{eq:6}
  \\
  E_{K}
  &= \hbar \omega_{K} \approx 2.4\,\mathrm{keV}\approx 4\times
  10^{-16}\,\mathrm{J}, \quad \frac{\Gamma_{\omega}}{\omega_{K}}
  \approx 10^{-2}. \label{eq:7}
\end{align}

Consequently, we conclude that for sulfur atoms, the maximum of the ionization cross section is located in the X-ray frequency range.

Let us now estimate the total radiation dose required to achieve a therapeutic effect. When a sulfur atom is ionized, the energy $E_{K}$ is absorbed. Thus, the total absorbed dose $D$ in Grays required for the ionization of $N_{S}$ sulfur atoms in a tumor of mass $M[\mathrm{kg}]$ is given by
\begin{align}
  D = \frac{E_{K}N_{S}}{M}. \label{eq:8}
\end{align}
From now on, we consider ionization as the main mechanism of radiation action on the sulfur atoms. Then
\begin{align}
  M = m_{S}N_{S0}, \label{eq:9}
\end{align}
where $m_{S}$ is the mass of the sulfur atom and $N_{S0}$ is the number of sulfur atoms before irradiation. As a result, after the irradiation, the number of ionized atoms
\begin{align}
  N_{S}
  &= N_{S0} D \frac{m_{S}}{E_{K}} = N_{S0} D
  \frac{32\cdot1.7\cdot10^{-27}}{2.4\cdot1.6\cdot10^{-16}}\nonumber
  \\
  &\approx 1.5\cdot10^{-10}N_{S0}D. \label{eq:10}
\end{align}
This means that if the total absorbed dose $D = 10\,\mathrm{Gy}$, the relative concentration of sulfur-containing molecules with destroyed bonds that is sufficient for the irreversible destruction of the tumor \cite{Kudrashov2004} equals
\begin{align}
 \frac{ N_{S}}{N_{S0}}\approx 10^{-9}.\label{eq:11}
\end{align}

Let us now determine the requirements for the X-ray source that can provide the required radiation dose. In a realistic situation, the quantity $N_{S}$ depends both on the spectral density of photons $I(\omega)$ from the source and on the surface area $\Sigma$ of the irradiated part of the tumor
\begin{align}
  N_{S} = N_{S0}\frac{1}{\Sigma}\int I(\omega)
  \sigma_{S}(\omega)d\omega. \label{eq:12}
\end{align}
Consequently, the total absorbed dose is given by
\begin{align}
  D = \frac{N_{S0}E_{K}}{M}\frac{1}{\Sigma}\int
  I(\omega)\sigma_{S}(\omega)d\omega.\label{eq:13}
\end{align}
As a result, the required therapeutic dose of radiation is strongly dependent on the source. For estimation, we assume that during irradiation the source emits $N_{\gamma}$ photons, which are uniformly distributed over the spectral interval $\Delta_{\omega}$, with a spectral density $I(\omega) = N_{\gamma} / \Delta_{\omega}$. Then, by performing the integral with the resonant part of the cross section Eq.~(\ref{eq:5}), one obtains the following estimate for the absorbed radiation dose
\begin{align}
  D = \frac{E_{K}N_{S0}}{M}\frac{\sigma_{0}}{\Sigma}\pi N_{\gamma}
  \frac{\Gamma_{\omega}}{\Delta_{\omega}}. \label{eq:14}
\end{align}

Usually, the source of radiation is characterized by the photon flux density, i.e., the number of photons $n_{\gamma}$ passing through a unit surface per unit time. Consequently, if the irradiation occurs over a time $T$ and the radiation penetrates an area $\Sigma_{0}$, the total number of photons is given by $N_{\gamma} = n_{\gamma} T \Sigma_{0}$. Thus, we find the expression for the absorbed radiation dose
\begin{align}
  D = \frac{E_{K}N_{S0}}{M}\pi\sigma_{0}\frac{\Sigma_{0}}{\Sigma}
  \frac{\Gamma_{\omega}}{\Delta_{\omega}} n_{\gamma} T. \label{eq:15}
\end{align}

From here, we find the condition on the photon flux density of the source
\begin{align}
  n_{\gamma} = D
  \frac{m_{S}}{\pi\sigma_{0}E_{K}T} \frac{\Sigma}{\Sigma_{0}}
  \frac{\Gamma_{\omega}}{\Delta_{\omega}}\,
  \left[\frac{\mathrm{quanta}}{\mathrm{cm}^{2}\times
  \mathrm{s}}\right]. \label{eq:16}
\end{align}

This quantity is optimal when $\Sigma = \Sigma_{0}$ and $\Delta_{\omega} = \Gamma_{\omega}$.

In order to deliver the above estimated absorbed radiation dose $D = 10\,\mathrm{Gy}$ during an irradiation time of $100\,\mathrm{s}$ over an area of $1\,\mathrm{cm}^{2}$, which is required for successful therapy, the source should provide the photon flux density
\begin{align}
  n_{\gamma}
  &\approx 10 \frac{32 \cdot 1.6 \cdot 10^{-27}}{3.14 \cdot
    9.2 \cdot 10^{-20} \cdot 2.4 \cdot 1.6 \cdot 10^{-16} \cdot 100}
    \left[\frac{\mathrm{quanta}}{\mathrm{cm}^2 \times
    \mathrm{s}}\right] \nonumber
  \\
  &\approx 5 \cdot 10^7 \left[\frac{\mathrm{quanta}}{\mathrm{cm}^2
    \times \mathrm{s}}\right]. \label{eq:17}
\end{align}

If the radiation source has a broad spectrum instead of a quasi-monochromatic one, then $\Delta_{\omega} \sim \omega_{K}$, and to deliver the same dose, a source with 100 times larger photon flux density is required.

In summary, the requirements for the radiation source for successful X-ray resonance therapy are:
\begin{enumerate}
\item The source should have tunable frequency, such that the maximum
  of intensity was located and the absorption $K$-edge of the sulphur
  atom $\hbar \omega_{K} = 2.4\,\mathrm{keV}$.
\item The source should emit the quasi-monochromatic X-ray radiation
  with the full width at half maximum $\Delta_{\omega} \approx 10^{-2}
  \omega_{K}$.
\item The spatial width of the photon beam should be around
  $1\,\mathrm{cm}^{2}$.
\item The source should emit approximately $\sim 10^{8}$ photons per
  second.
\item The whole experimental setup should be compact, thus allowing its
  location at the ordinary hospitals.
\end{enumerate}

In the next section, we will demonstrate that a radiation source based on PXR satisfies all the above requirements.

\section{PXR characteristics}
\label{sec:qualitative}

Parametric X-ray radiation (PXR) is emitted when the electron bunch moves through the crystal. This mechanism of X-ray generation has been investigated in detail both theoretically and experimentally \cite{PXR_Book_Feranchuk}. Therefore, we will not describe the theory of this effect, but will focus only on the characteristics of PXR that are essential for the present problem.

\begin{figure*}
\centering
\includegraphics[scale=0.7]{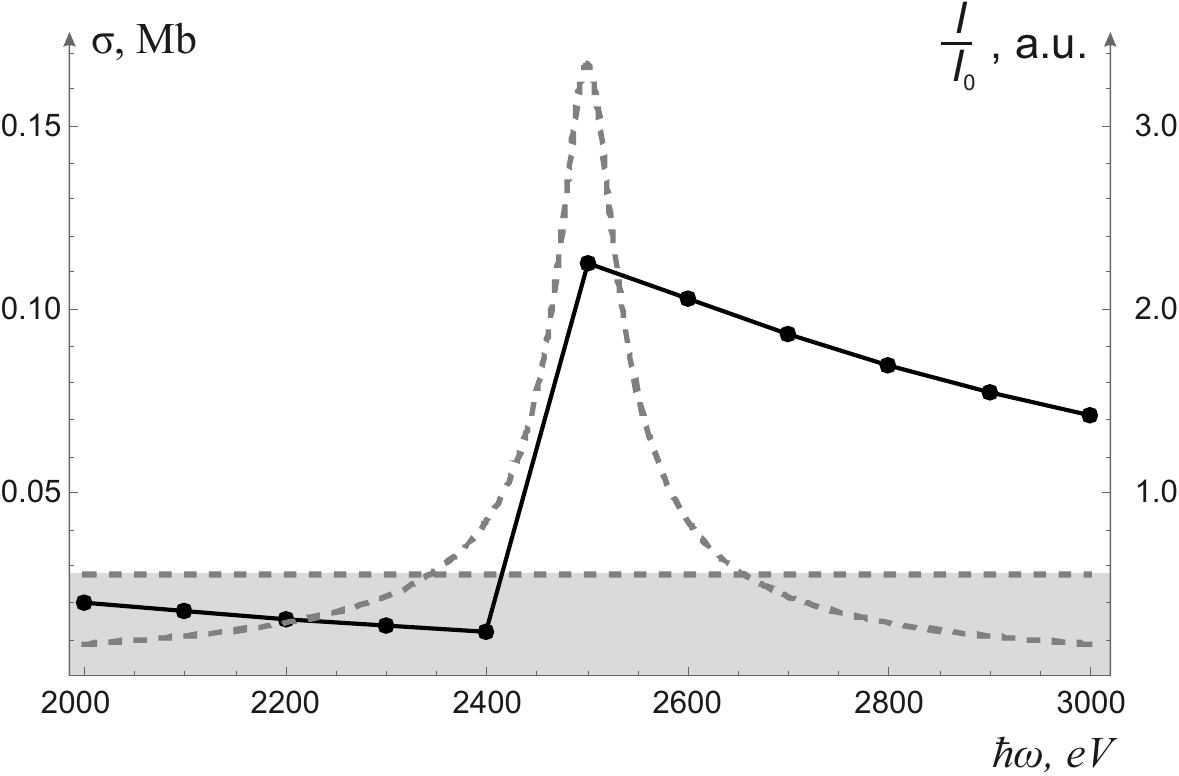}
\caption{Spectra photo-ionisation of S (solid line) and PXR (dash line)}
\label{fig:1}
\end{figure*}

When an electron moves in the crystal with velocity $\vec v$ at an angle $\theta_B$ to the crystallographic planes, part of its electromagnetic self-field is reflected from these planes and exits the crystal as a photon beam directed at an angle $2\theta_B$ relative to the electron velocity
(Fig.~\ref{fig:1}).

\begin{figure*}
\centering
\includegraphics[scale=0.7]{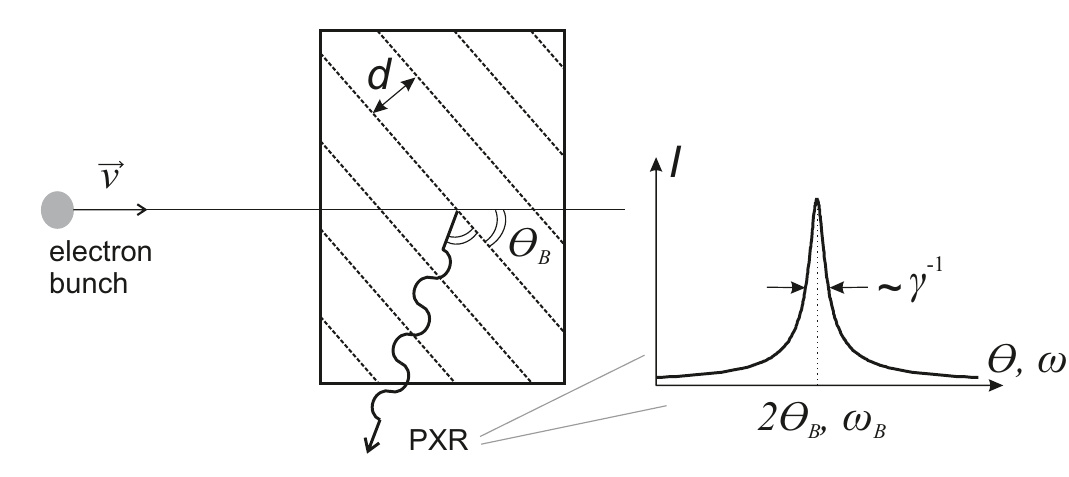}
\caption{Scheme of the PXR generation}
\label{fig:2}
\end{figure*}

The characteristic frequency of these photons depends on the interplane distance $d$, corresponds to the X-ray range, and is defined by the formula
\begin{align}
  \omega_B = \frac{\pi c}{d \sin \theta_B}, \label{eq:18}
\end{align}
where $c$ is the speed of light.

The spectral and angular distributions of the emitted photons have the form of sharp peaks with widths $\Delta \omega$ and $\Delta \theta$, respectively, depending on the electron energy (Fig.~\ref{fig:2}):
\begin{eqnarray}
\label{7}
\frac{\Delta\omega}{\omega_B}\approx \Delta \theta \approx \gamma^{-1};  \ \gamma = \frac{E}{mc^2}.
\end{eqnarray}

Importantly, the frequency of the emitted photons can be smoothly tuned simply by rotating the crystal and arranging it at such an angle to the electron velocity that the frequency of the PXR peak coincides with $\omega_K$:
\begin{eqnarray}
\label{9}
\sin \theta_B = \frac{\pi c}{d \omega_K}.
\end{eqnarray}

If, for example, a Si crystal is used for PXR generation, the angle between the electron velocity and the $(111)$ plane should be equal to
\begin{eqnarray}
\label{10}
\theta_B = \frac{\pi c \sqrt{3}}{a \omega_K} \approx 51.26^o .
\end{eqnarray}
Here $a \approx 5.43  \text{\AA} $ is the length of the crystal unit cell.

In this case, the PXR spectral distribution has a maximum at the K-edge of sulfur (Fig.~\ref{fig:2}).

An important characteristic of the radiation source is the total number of photons generated in the peak per unit time. If the crystallographic plane with indices $(hkl)$ is used for PXR generation, then the number of photons emitted per second can be calculated by the formula \cite{PXR_Book_Feranchuk}:
\begin{eqnarray}
\label{11}
N_{ph} \approx \frac{\alpha \cos 2 \theta_B(1+ \cos^2 2 \theta_B) }
{4 \sin^2\theta_B}\frac{J}{e}\frac{|\chi_{hkl}(\omega_B)|^2 }{|\chi_0"(\omega_B)|}\ln(2\gamma).
\end{eqnarray}
Here $J$ is the average current of the electron beam, $e$ is the electron charge, $\alpha$ is the fine structure constant, and $\chi_{hkl}$ and $\chi_0 = \chi_0' + i\chi_0"$ are the components of the crystal polarizability \cite{benediktovich2013theoretical}.

Suppose the (111) plane of a Si crystal is used for PXR generation with frequency $\omega_B = \omega_K$. In this case, $|\chi_{111}(\omega_K)| = 8.7 \times 10^{-5}$ and $|\chi_0"(\omega_K)|= 3.6 \times 10^{-5}$ \cite{StepanovXrayWebServer}. Then, for a beam of electrons with energy $E = 25$ MeV and average current 1 mA, typical for medical accelerators \cite{Accelerator}, formula (\ref{11}) gives the following estimate for the photon intensity:
\begin{eqnarray}
\label{12}
N_{ph} \approx 3 \ 10^{11} \ \frac{ph}{s},
\end{eqnarray}
and the X-ray beam is concentrated in rather small spectral and angular ranges:
\begin{eqnarray}
\label{7a}
\frac{\Delta\omega}{\omega_B}\approx \Delta \theta \approx 10^{-2}.
\end{eqnarray}

\section{Efficiency of the irradiation on the basis of PXR}

Let us consider the conditions when the suggested resonant therapy could be effective for the tumor irradiation.

First of all, it should be taken into account that rather soft X-rays are used, which penetrate the medium to a depth of $L \sim 1 \ \mathrm{mm}$ depending on the tumor composition. Therefore, selective irradiation can be useful for the therapy of surface tumors. The photon beam can be delivered to the irradiation spot using the same equipment as suggested for an X-ray scalpel \cite{gutman2007x} or with an X-ray lens \cite{dudchik1999microcapillary}.

Let the concentration of sulfur atoms in the tumor molecules be $N_0 \ \mathrm{cm}^{-3}$. Then the number of ionized sulfur atoms $N_S$ that appear in the tumor when $N_{ph}$ photons pass through the medium per unit time depends on the photo-ionization cross-section and can be estimated by formula:
\begin{eqnarray}
\label{13}
N_{S} \approx \sigma_0N_0 L N_{ph} \ [s^{-1}],
\end{eqnarray}
and the concentration of ionized sulfur atoms $N_i$ equals
\begin{eqnarray}
\label{14}
N_i  = \frac{N_S}{L \Sigma}t \approx \frac{\sigma_0    N_{ph} t}{\Sigma} N_0
\end{eqnarray}
Here $t$ is the irradiation time; $\Sigma = R^2 \gamma^{-2}$ is the surface area covered by the PXR photon beam; $R$ is the distance between the X-ray source and the tumor tissue. If the electron energy is $E = 25$ MeV and the distance $R\approx 10^2 \ \mathrm{cm}$, then $\Sigma\approx 1\ \mathrm{cm}^2$. The maximum value of the photo-ionization cross-section at the photon frequency $\omega_K$ is $\sigma_0 \approx 1.1 \times 10^{-19}\ \mathrm{cm}^2$. Then, during an irradiation time $t = 600\ \mathrm{s}$, the concentration of ionized sulfur atoms is
\begin{eqnarray}
\label{15}
N_i  \approx 3 \times 10^{-4}  N_0
\end{eqnarray}

This value is much larger than the critical concentration (\ref{11}), $N^{cr}_i (D)\sim 10^{-7}$, which is necessary  for destruction of the tumor.

It is important to estimate the total dose absorbed by the patient during such a session of irradiation. It can be calculated by the following formula:
\begin{eqnarray}
\label{16}
D = \frac{\hbar \omega_K  N_{ph} t}{\rho V} \ Gy.
\end{eqnarray}
Here, the photon energy $\hbar \omega_K \approx 4 \times 10^{-16}\ \mathrm{J}$, the average density $\rho\approx 4 \times 10^3\ \mathrm{kg}/\mathrm{m}^3$, and $V\ (\mathrm{m}^3)$ is the volume of the medium where the absorbed dose is distributed. It depends on the heat conductivity of the tumor medium:
\begin{eqnarray}
\label{17}
V \approx 8 (a t) ^{3/2},
\end{eqnarray}
where $a \approx 10^{-6}\ \mathrm{m}^2/\mathrm{s}$ is the coefficient of heat conductivity. When these parameters are used, the total absorbed dose equals:
\begin{eqnarray}
\label{18}
D \approx 1.3 \ Gy
\end{eqnarray}

If a conventional X-ray tube is used for the irradiation, the same effect is obtained, but the total absorbed dose is two orders of magnitude larger \cite{balanov2024coherentinteractionsfreeelectrons}.
 
\section{Conclusions}

In conclusion, the quantitative estimates presented in this research demonstrate that PXR provides a highly effective and selective platform for the resonance therapy of superficial sulfur-containing tumor tissues. By precisely tuning the PXR frequency to the K-shell ionization energy of sulfur atoms ($\hbar\omega_K \approx 2.4$ keV) through simple crystal rotation, the proposed method successfully exploits the significant biochemical disparity in sulfur concentration—which is approximately three times higher in specific malignant tissues than in healthy ones. The results indicate that the quasi-monochromatic nature and small angular divergence of the PXR beam allow for a therapeutic effect with a total absorbed dose of only $1.3$ Gy, a value that is two orders of magnitude lower than the dose required by conventional X-ray tubes. Furthermore, unlike synchrotron-based methods that necessitate large-scale $1$ GeV accelerators, PXR-based sources can operate effectively at electron energies of $20$--$40$ MeV, which are standard for modern medical accelerators. This technical advantage allows for the development of compact experimental setups suitable for standard hospital environments. While the relatively soft X-rays limit penetration to a depth of approximately 1 mm, the estimated concentration of ionized sulfur atoms ($3 \times 10^{-4} N_0$) remains well above the critical threshold for irreversible tumor destruction, confirming the method's potency for targeted superficial treatment. Consequently, this approach not only minimizes the collateral damage to healthy cells but also offers a more efficient and accessible paradigm for clinical radiation oncology through the integration of existing medical technologies like X-ray scalpels and lenses.

\section{Acknowledgments}

The authors dedicate this work to the memory of Professor Oleg Iosifovich Shadyro.

The authors are grateful to Prof. A.P.Ulyanenkov  for the support of this work.

This work was granted by Hue University under project number DHH2025-18-05.

\bibliography{medical_pxrr}

@article{sandoghdar2024essay,
  title = {Essay: Exploring the Physics of Basic Medical Research},
  author = {Sandoghdar, Vahid},
  journal = {Phys. Rev. Lett.},
  volume = {132},
  issue = {9},
  pages = {090001},
  numpages = {7},
  year = {2024},
  month = {Feb},
  publisher = {American Physical Society},
  doi = {10.1103/PhysRevLett.132.090001},
  url = {https://link.aps.org/doi/10.1103/PhysRevLett.132.090001}
}

@article{balanov2024coherentinteractionsfreeelectrons,
author = {Amnon Balanov and Alexey Gorlach and Vladimir Baryshevsky and Ilya Feranchuk and Hideo Nitta and Yasushi Hayakawa and Alexander Shchagin and Yuichi Takabayashi and Yaron Danon and Liang Jie Wong and Ido Kaminer},
journal = {Adv. Opt. Photon.},
number = {4},
pages = {726--788},
publisher = {Optica Publishing Group},
title = {Coherent interactions of free electrons and matter: toward tunable compact x-ray sources},
volume = {17},
month = {Dec},
year = {2025},
url = {https://opg.optica.org/aop/abstract.cfm?URI=aop-17-4-726},
doi = {10.1364/AOP.559742},
}

@article{labriet2018significant,
  title={Significant dose reduction using synchrotron radiation computed tomography: first clinical case and application to high resolution CT exams},
  author={Labriet, H and Nemoz, C and Renier, M and Berkvens, P and Brochard, T and Cassagne, R and Elleaume, H{\'e}l{\`e}ne and Est{\`e}ve, Fran{\c{c}}ois and Verry, Camille and Balosso, Jacques and others},
  journal={Scientific reports},
  volume={8},
  number={1},
  pages={12491},
  year={2018},
  publisher={Nature Publishing Group UK London},
  doi = {10.1038/s41598-018-30902-y},
  url = {https://doi.org/10.1038/s41598-018-30902-y}
}

@Inbook{Podgorsak2014,
author="Podgor{\v{s}}ak, Ervin B.",
title="Particle Accelerators in Medicine",
bookTitle="Compendium to Radiation Physics for Medical Physicists: 300 Problems and Solutions",
year="2014",
publisher="Springer Berlin Heidelberg",
address="Berlin, Heidelberg",
pages="1041--1099",
isbn="978-3-642-20186-8",
doi="10.1007/978-3-642-20186-8_14",
url="https://doi.org/10.1007/978-3-642-20186-8_14"
}

@book{Kudrashov2004,
Author = {Yu.B.Kudryashov and V.K.Mazurik and M.F.Lomanov},
 Title = {Radiation biophysics (ionizing radiation)},
  Publisher = {Fizmatlit},
  Year={2004},
  note={[in russian]}
}

@book{kudriashov2008radiation,
  title={Radiation Biophysics (Ionizing Radiations)},
  author={Yurii B. Kudryashov},
  year={2008},
  publisher={Nova Publishers},
  url = {https://catalog.nlm.nih.gov/discovery/fulldisplay?docid=alma9914738433406676&context=L&vid=01NLM_INST:01NLM_INST&lang=en&search_scope=MyInstitution&adaptor=Local%20Search%20Engine&tab=LibraryCatalog&query=mesh,exact,Radiation,%20Ionizing,AND&mode=advanced&offset=150}
}

@article{Kuzin1991,
  Title={The action of atomic radiation in low doses on the biota},
  author={Kuzin, AM},
  journal={Radiobiologiya},
  volume={31},
  number={2},
  pages={175--179},
  year={1991},
  note={[in russian]}
}

@article{gamcsik2012glutathione,
  title={Glutathione levels in human tumors},
  author={Gamcsik, Michael P and Kasibhatla, Mohit S and Teeter, Stephanie D and Colvin, O Michael},
  journal={Biomarkers},
  volume={17},
  number={8},
  pages={671--691},
  year={2012},
  publisher={Taylor \& Francis},
  doi = {10.3109/1354750X.2012.715672},
  url = {https://doi.org/10.3109/1354750X.2012.715672}
}

@article{huang2001mechanism,
  title={Mechanism and significance of increased glutathione level in human hepatocellular carcinoma and liver regeneration},
  author={Huang, Zong-Zhi and Chen, Changjin and Zeng, Zhaohui and Yang, Heping and Oh, Jenny and Chen, Lixin and C. Lu, Shelly},
  journal={The FASEB Journal},
  volume={15},
  number={1},
  pages={19--21},
  year={2001},
  publisher={Wiley Online Library},
  doi = {10.1096/fj.00-0445fje},
  url = {https://doi.org/10.1096/fj.00-0445fje}
}

@article{CELLARIER2003489,
title = "Methionine dependency and cancer treatment",
journal = "Cancer Treatment Reviews",
volume = "29",
number = "6",
pages = "489 - 499",
year = "2003",
issn = "0305-7372",
doi = "10.1016/S0305-7372(03)00118-X",
url = "http://www.sciencedirect.com/science/article/pii/S030573720300118X",
author = "E Cellarier and X Durando and M.P Vasson and M.C Farges and A Demiden and J.C Maurizis and J.C Madelmont and P Chollet"
}

@article{czapla2013chemical,
  title={Chemical species of sulfur in prostate cancer cells studied by XANES spectroscopy},
  author={Czapla, Joanna and Kwiatek, Wojciech M and Lekki, Janusz and Duli{\'n}ska-Litewka, Joanna and Steininger, Ralph and G{\"o}ttlicher, J{\"o}rg},
  journal={Radiation Physics and Chemistry},
  volume={93},
  pages={154--159},
  year={2013},
  publisher={Elsevier},
  doi = {10.1016/j.radphyschem.2013.05.021},
  url = {https://doi.org/10.1016/j.radphyschem.2013.05.021}
}

@book{Landau2002,
  title={Course of Theoretical Physics. Volume III. Quantum Mechanics: Non-relativistic Theory},
  author={L. D. Landau and E. M. Lifshitz},
  year={1977},
  publisher={Pergamon Press}
}

@book{mirkin1964handbook,
  title={Handbook of X-ray analysis of polycrystalline materials},
  author={Mirkin, Lev Iosifovich  },
  year={1964},
  publisher={Springer},
  url = {https://link.springer.com/book/9781468460629}
}

@book{PXR_Book_Feranchuk,
  Author = {Vladimir G. Baryshevsky and Ilya D. Feranchuk and Alexander P. Ulyanenkov},
  Title = {Parametric X-Ray Radiation in Crystals: Theory, Experiment and Applications (Springer Tracts in Modern Physics)},
  Publisher = {Springer},
  Year = {2006},
  ISBN = {3540269053},
  URL = {http://www.springer.com/de/book/9783540269052}
}

@book{benediktovich2013theoretical,
  title={Theoretical Concepts of X-Ray Nanoscale Analysis: Theory and Applications},
  author={Benediktovich, A. and Feranchuk, I. and Ulyanenkov, A.},
  isbn={9783642381768},
  lccn={2013942118},
  series={Springer Series in Materials Science},
  url={http://www.springer.com/gp/book/9783642381768},
  year={2013},
  publisher={Springer Berlin Heidelberg}
}

@misc{StepanovXrayWebServer,
  title = {X-ray dynamical diffraction web server},
  howpublished = {\url{http://x-server.gmca.aps.anl.gov/}},
  note = {Accessed: 07-03-2019},
  author = {Stepanov, S. A.}
}

@article{Accelerator,
  title={Nuclear-physical technologies in medicine},
  author={Chernyaev, AP},
  journal={Physics of elementary particles and atomic nuclei},
  volume={43},
  number={2},
  pages={499--518},
  year={2012},
  note={[in russian]}
}

@article{gutman2007x,
  title={X-ray scalpel—a new device for targeted x-ray brachytherapy and stereotactic radiosurgery},
  author={Gutman, George and Strumban, Emil and Sozontov, Evgeny and Jenrow, Kenneth},
  journal={Physics in Medicine \& Biology},
  volume={52},
  number={6},
  pages={1757},
  year={2007},
  publisher={IOP Publishing},
  doi = {10.1088/0031-9155/52/6/015},
  url = {https://iopscience.iop.org/article/10.1088/0031-9155/52/6/015}
}

@article{dudchik1999microcapillary,
  title={A microcapillary lens for X-rays},
  author={Dudchik, Yu I and Kolchevsky, NN},
  journal={Nuclear Instruments and Methods in Physics Research Section A: Accelerators, Spectrometers, Detectors and Associated Equipment},
  volume={421},
  number={1-2},
  pages={361--364},
  year={1999},
  publisher={Elsevier},
  doi = {10.1016/S0168-9002(98)01269-8},
  url = {https://doi.org/10.1016/S0168-9002(98)01269-8}
}
\end{document}